\documentclass[10pt]{article}
\usepackage{graphicx}
\usepackage{amsmath}
\usepackage{amssymb}
\usepackage{caption2}
\begin{document}
\begin{center}
{\large {\bf \sc{  Analysis of the $X(4475/4710)$, $X(4500/4700)$,
 $Z_{c}(4600)$, $Z_{\bar{c}\bar{s}}(4600)$ and related tetraquark states with the QCD sum rules }}} \\[2mm]
Zhi-Gang  Wang \footnote{E-mail: zgwang@aliyun.com.  }     \\
 Department of Physics, North China Electric Power University, Baoding 071003, P. R. China
\end{center}

\begin{abstract}
In this work,  we introduce an explicit P-wave to construct the diquarks $[qc]_{\widehat{V}}$, then construct  the local four-quark  currents to explore  the  hidden-charm  tetraquark states with the  $J^{PC}=0^{++}$, $1^{+-}$ and $2^{++}$ in the framework of the  QCD sum rules at length.  Our calculations indicate that the  light-flavor $SU(3)$ breaking effects on the tetraquark masses are tiny. The  predictions   support
assigning the  $X(4475)$ and $X(4500)$ as the
$[uc]_{\widehat{V}}[\overline{uc}]_{\widehat{V}}-[dc]_{\widehat{V}}[\overline{dc}]_{\widehat{V}}$  and $[sc]_{\widehat{V}}[\overline{sc}]_{\widehat{V}}$ tetraquark states with the $J^{PC}=0^{++}$  respectively, and  assigning the $Z_{c}(4600)$ and $Z_{\bar{c}\bar{s}}(4600)$ as the
   $[uc]_{\widehat{V}}[\overline{dc}]_{\widehat{V}}$ and $[qc]_{\widehat{V}}[\overline{sc}]_{\widehat{V}}$
 tetraquark states with the
$J^{PC}=1^{+-}$ respectively. On the other hand, there is no room for the $X(4710)$ and $X(4700)$. Combined with  previous works, the $X(4475)$, $X(4500)$, $Z_{c}(4600)$ and $Z_{\bar{c}\bar{s}}(4600)$  might have other important Fock components besides  the $\widehat{V}\widehat{V}$ type components.
 \end{abstract}


\section{Introduction}
In 2016, the LHCb collaboration performed the first full amplitude analysis of the decays $B^+\to J/\psi \phi K^+$  with a data sample of
 $3\,{\rm fb}^{-1}$ of the $pp$ collision data  at $\sqrt{s}=7$ and $8\, \rm{TeV}$ \cite{LHCb-16061,LHCb-16062}.
And they observed four $J/\psi\phi$ structures:  two old particles $X(4140)$ and $X(4274)$ and  two new particles $X(4500)$ and $X(4700)$. The statistical significances of the $X(4140)$, $X(4274)$, $X(4500)$ and $X(4700)$ are
$8.4\sigma$, $6.0\sigma$, $6.1\sigma$ and $5.6\sigma$, respectively. While
the statistical significances of their  quantum numbers  $J^{PC}=1^{++}$, $1^{++}$, $0^{++}$ and $0^{++}$ are  $5.7\sigma$, $5.8\sigma$, $4.0\sigma$ and $4.5\sigma$, respectively \cite{LHCb-16061,LHCb-16062}.
The LHCb collaboration determined the $J^{PC}$ of
the $X(4140)$  to be $1^{++}$, thus  ruling out the $0^{++}/2^{++}$
$D_s^{*+}D_s^{*-}$ molecule assignments.

In 2021, the LHCb collaboration performed an improved full amplitude analysis of the exclusive $B^+\to J/\psi \phi K^+$ decays  using the $pp$ collision data (6 times larger signal yield than previously analyzed) corresponding to a total integrated luminosity of $9{\rm fb}^{-1}$  at $\sqrt{s}=7$, $8$ and $13$ $\rm TeV$ \cite{LHCb-X4685}. They  observed the $Z_{cs}(4000)$ with the $J^P=1^+$ in the $J/\psi K^+$ mass spectrum with the statistical significance of $15\sigma$, and  the $X(4685)$  ($X(4630)$) in the $J/\psi \phi$  mass spectrum with the $J^P=1^+$ ($1^-$) with the statistical significance of $15\sigma$  ($5.5\sigma$).  Furthermore, they confirmed the old particles $X(4140)$, $X(4274)$, $X(4500)$ and $X(4700)$. The measured  Breit-Wigner masses and widths are,
\begin{flalign}
 & X(4140) : M = 4118 \pm 11{}_{-36}^{+19} \mbox{ MeV}\, , \, \Gamma = 162\pm 21_{-49}^{+24} \mbox{ MeV} \, , \nonumber\\
 & X(4274) : M = 4294\pm4 _{-6}^{+3} \mbox{ MeV}\, , \, \Gamma = 53\pm5\pm5 \mbox{ MeV} \, ,\nonumber \\
 & X(4685) : M = 4684\pm 7_{-16}^{+13} \mbox{ MeV}\, , \, \Gamma = 126\pm15 _{-41}^{+37} \mbox{ MeV} \, ,\nonumber \\
 & X(4500) : M = 4474\pm3 \pm3 \mbox{ MeV} \, ,\, \Gamma = 77\pm 6 _{-8}^{+10} \mbox{ MeV} \, , \nonumber\\
 & X(4700) : M = 4694\pm 4 _{-3}^{+16} \mbox{ MeV} \, ,\, \Gamma = 87\pm8 _{-6}^{+16} \mbox{ MeV} \, ,\nonumber\\
 & X(4630) : M = 4626\pm 16_{-110}^{+18} \mbox{ MeV} \, ,\, \Gamma = 174\pm27 _{-73}^{+134} \mbox{ MeV} \, .
\end{flalign}
If we adopt the scenario of the color $\bar{\mathbf{3}}\mathbf{3}$-type tetraquark states and consult  the predictions based on the QCD sum rules, the $X(4140)$ and $X(4274)$ lie at the 1S region, the $X(4500)$, $X(4685)$ and $X(4700)$ lie at the 2S region, the $X(4630)$ lies at the 1P region, etc  \cite{WZG-HC-spectrum-NPB,WZG-NPB-cscs,WZG-review}.

In 2024, the LHCb collaboration performed the first full amplitude analysis of the decays $B^+ \to \psi(2S) K^+ \pi^+ \pi^-$  using the $pp$ collision data corresponding to an integrated luminosity of $9\,\text{fb}^{-1}$, and they developed   an amplitude model with 53 components  comprising 11 hidden-charm exotic states: $Z_c(4055)$, $Z_c(4200)$, $Z_c(4430)$, $X(4475)$, $X(4710)$, $X(4650)$, $X(4800)$, $Z_{\bar{c}\bar{s}}(4000)$, $Z_{\bar{c}\bar{s}}(4600)$, $Z_{\bar{c}\bar{s}}(4900)$, $Z_{\bar{c}\bar{s}}(5200)$ \cite{LHCb-X4500-ccqq}.

They confirmed the $Z_c(4200)$ and $Z_c(4430)$ in the $\psi(2S) \pi^+$ mass spectrum, and  determined the spin-parity of the $Z_c(4200)$ to be $1^+$ for the first time with a significance exceeding $5\sigma$, and observed the $Z_{\bar{c}\bar{s}}(4600)$ and $Z_{\bar{c}\bar{s}}(4900)$ in the $\psi(2S)K^*(892)$ mass spectrum. The measured  Breit-Wigner masses and widths are \cite{LHCb-X4500-ccqq},
\begin{flalign}
  & Z_c(4200) : M = 4257 \pm  11 \pm 17 \mbox{ MeV} \, ,\, \Gamma = 308 \pm 20 \pm 32 \mbox{ MeV} \, , \nonumber \\
 & Z_c(4430) : M = 4468 \pm 21 \pm 80 \mbox{ MeV} \, ,\, \Gamma = 251 \pm 42 \pm 82 \mbox{ MeV} \, , \nonumber \\
  & Z_{\bar{c}\bar{s}}(4600) : M = 4578  \pm  10  \pm  18 \mbox{ MeV} \, ,\, \Gamma = 133  \pm  28  \pm  69\mbox{ MeV} \, , \nonumber \\
& Z_{\bar{c}\bar{s}}(4900) : M = 4925  \pm  22  \pm  47 \mbox{ MeV} \, ,\, \Gamma = 255  \pm  55  \pm  127\mbox{ MeV} \, .
\end{flalign}

 In 2019, the LHCb collaboration performed an angular analysis of the  decays $B^0\to J/\psi K^+\pi^-$ using the $pp$ collision data corresponding to an integrated luminosity of $3\,\rm{fb}^{-1}$, examined the $m(J/\psi \pi^-)$ versus the $m(K^+\pi^-)$ plane, and observed two  structures in the vicinity of  the energies $m(J/\psi \pi^-)=4200 \,\rm{MeV}$ and $4600\,\rm{MeV}$, respectively \cite{LHCb-Z4600}. However, the $Z_c(4600)$ has not been confirmed yet.

Again,  we adopt the scenario of the color $\bar{\mathbf{3}}\mathbf{3}$-type tetraquark states and consult  the predictions based on the QCD sum rules, the $Z_c(4430)$, $Z_c(4600)$ and  $Z_{\bar{c}\bar{s}}(4600)$ lie at the 2S region, the $Z_{\bar{c}\bar{s}}(4900)$ lies  at the 3S region, while there is no room for the $Z_c(4200)$   \cite{WZG-review,WangZG-Z4430-CTP,ChenHX-Z4600-A,WangZG-axial-Z4600}.

 In addition, the LHCb collaboration  observed that the $\psi(2S) \pi^+ \pi^-$ mass spectrum are dominated by the $X^0 \to \psi(2S) \rho^0(770)$ decays with $X^0=X(4475)$, $X(4650)$, $X(4710)$ and $X(4800)$ \cite{LHCb-X4500-ccqq}, which are similar  to the previously
observed $J/\psi\phi$ resonances $X(4500)$, $X(4685)$, $X(4700)$ and $X(4630)$, respectively \cite{LHCb-X4685}.
 The spin-parity  of the $X(4630)$ have not been unambiguously determined yet,  the assignment $J^P =1^-$ is favored over $J^P =2^-$ with a significance of $3\sigma$ and other assignments are disfavored by more than $5\sigma$ \cite{LHCb-X4685}.
 As the strong decays conserve isospin, we can sort out those states according to the isospins of the final states $\psi(2S)\rho^0(770)$
and $J/\psi\phi(1020)$,
  \begin{flalign}
 & (I,I_3)=(1,0) : X(4475)\, , \, X(4650)\, , \, X(4710)\, , \, X(4800) \, , \nonumber\\
 & (I,I_3)=(0,0) : X(4500)\, , \, X(4685)\, , \, X(4700)\, , \, X(4630) \, .
\end{flalign}
The measured  Breit-Wigner masses and widths are \cite{LHCb-X4500-ccqq},
\begin{flalign}
 & X(4475) : M = 4475 \pm 7 \pm 12 \mbox{ MeV}\, , \, \Gamma = 231 \pm19 \pm 32 \mbox{ MeV} \, , \nonumber\\
 & X(4650) : M = 4653 \pm 14 \pm 27 \mbox{ MeV}\, , \, \Gamma = 227 \pm 26 \pm 22 \mbox{ MeV} \, ,\nonumber \\
 & X(4710) : M = 4710 \pm 4 \pm 5 \mbox{ MeV} \, ,\, \Gamma = 64 \pm9 \pm 10 \mbox{ MeV} \, , \nonumber \\
 & X(4800) : M = 4785\pm 37 \pm 119 \mbox{ MeV} \, ,\, \Gamma = 457
\pm 93 \pm 157 \mbox{ MeV} \, .
 \end{flalign}
A possible  explanation is that those states are genuinely different states, however, for example, if the $X(4475)$ state is
the $c\bar{c}(u\bar{u} -d\bar{d})$ isospin partner of the $X(4500)$ which interpreted as the $c\bar{c}s\bar{s}$ state, we would generally
expect a larger mass difference of $M_{X(4500)}-M_{X(4475)}\approx 200\,\rm{MeV}$ rather than several $\rm{MeV}$. It is interesting to explore the light-flavor $SU(3)$ breaking effects.

On the other hand, we should bear in mind that  the observation of the $Z_c(4055)$, $X(4800)$ and  $Z_{\bar{c}\bar{s}}(5200)$ should not be considered as confirmations of specific states but
rather effective descriptions of the generic $\psi(2S)\pi^+$, $\psi(2S)\rho^0(770)$ and $\psi(2S)[K^+\pi^-]_S$ contributions respectively with the  $J^P=1^-$ \cite{LHCb-X4500-ccqq}.

Also in 2024, the LHCb collaboration  accomplished  the first investigation  of the $J/\psi\phi$ production in diffractive processes in the $pp$ collisions, which  is based on a data-set recorded at  $\sqrt{s}=13\,\mathrm{TeV}$ corresponding to an integrated luminosity of $5\,\mathrm{fb}^{-1}$ \cite{LHCb-X4500-X4274}.
The data  are consistent with a resonant model including several resonant states observed previously  in the $B^+\to J/\psi\phi K^+$ decays.
The $X(4500)$ and $X(4274)$ were observed with significances over $5\,\sigma$ and $4\,\sigma$, respectively.

Now we reach a short summary. The $X(4140)$, $X(4274)$, $X(4500)$, $X(4630)$, $X(4685)$ and $X(4700)$ observed in the $J/\psi\phi$  mass spectrum have the symbolic  valence quarks $c\bar{c}s\bar{s}$, isospin $(I,I_3)=(0,0)$ and $J^{PC}=0^{++}$, $1^{++}$, $2^{++}$ for the S-wave systems and $0^{-+}$, $1^{-+}$, $2^{-+}$, $3^{-+}$ for the P-wave systems.
The  $X(4475)$, $X(4650)$, $X(4710)$ and $X(4800)$  observed in the $\psi(2S)\rho^0(770)$  mass spectrum have the symbolic  valence quarks $c\bar{c}(u\bar{u}-d\bar{d})$, isospin $(I,I_3)=(1,0)$ and $J^{PC}=0^{++}$, $1^{++}$, $2^{++}$ for the S-wave systems and $0^{-+}$, $1^{-+}$, $2^{-+}$, $3^{-+}$ for the P-wave systems.
The $Z_{\bar{c}\bar{s}}(4600)$ and $Z_{\bar{c}\bar{s}}(4900)$ ($Z_{\bar{c}\bar{s}}(4000)$)  observed in the $\psi(2S)K^*(892)$ ($\psi(2S)K$) mass spectrum have the symbolic  valence quarks $c\bar{c}u\bar{s}$, isospin $(I,I_3)=(\frac{1}{2},\frac{1}{2})$ and $J^{PC}=1^{+-}$ for the S-wave systems.

In the following,  we would like to see from the perspective of the tetraquark picture for the $X$, $Y$ and $Z$ states and resort to the QCD sum rules \cite{WZG-review}.
We often  take the diquarks as the basic valence constituents  to study   the tetraquark states.   The diquarks  $\varepsilon^{ijk}q^{T}_j C\Gamma q^{\prime}_k$  have  five  spinor  structures, where  $C\Gamma=C\gamma_5$, $C$, $C\gamma_\mu \gamma_5$,  $C\gamma_\mu $ and $C\sigma_{\mu\nu}$ (or $C\sigma_{\mu\nu}\gamma_5$) for the scalar ($S$), pseudoscalar ($P$), vector ($V$), axialvector ($A$)  and  tensor ($T$) diquarks, respectively,  the $i$, $j$, $k$ are color indexes. And the $T$-diquarks have both the $J^P=1^+$ ($\widetilde{A}$) and $1^-$ ($\widetilde{V}$) components.

In Ref.\cite{WangHuangTao-3900}, we  explored  the energy scale dependence of the QCD sum rules for the $X$, $Y$ and $Z$ states for the first time. Later,  we
suggested  an energy scale formula,
\begin{eqnarray}\label{ESF}
\mu&=&\sqrt{M^2_{X/Y/Z}-(2{\mathbb{M}}_Q)^2} \, ,
 \end{eqnarray}
 via introducing the effective heavy quark masses ${\mathbb{M}}_Q$ to obtain the  suitable energy scales of the QCD spectral densities for  the hidden-charm (hidden-bottom) tetraquark states \cite{WangTetraquarkCTP,Wang-tetra-formula,Wang-Huang-NPA-2014}. The formula  can magnify the ground state contributions substantially
 and improve  convergence   of the operator product expansion substantially. If there exist valence $s$-quarks,  we  modify the energy scale formula,
\begin{eqnarray}\label{MESF}
\mu&=&\sqrt{M^2_{X/Y/Z}-(2{\mathbb{M}}_Q)^2}-\kappa\, {\mathbb{M}}_s \, ,
 \end{eqnarray}
to account for the light-flavor $SU(3)$ breaking effects via introducing  the effective $s$-quark mass ${\mathbb{M}}_s$, where the $\kappa$ is the valence $s$-quark's number.
The modified energy scale formula plays an important role in exploring the
light-flavor $SU(3)$ breaking effects in the multiquark states, especially those $X$ states \cite{WZG-HC-spectrum-NPB,WZG-NPB-cscs,
WZG-IJMPA-Pcs4469,WZG-mole-IJMPA,WZG-XQ-mole-penta,WZG-XQ-mole-EPJA,
WZG-tetra-psedo-NPB,WZG-Zcs3985-4110,WZG-NPB-cucd,
WZG-HC-spectrum-PRD}.

We should bear in mind that there exist other theoretical schemes (irrespective  the tetraquark picture \cite{X4140-X4700-tetra-ChenHX,X4140-X4700-tetra-LuQF,X4140-X4700-tetra-HuangF,X4140-X4700-tetra-WuJ,
X4140-X4700-tetra-Ferretti-SB,X4140-X4700-tetra-HuangHX},   molecule picture \cite{X4140-X4700-mole-PengFZ,X4140-X4700-mole-WangE,X4140-X4700-mole-GuoFK} or charmonium picture \cite{X4140-X4700-cc-ZhongXH,X4140-X4700-cc-ChenDY,X4140-X4700-cc-LiuXH,cc-Godfrey})  and possible assignments for those $X$ states. All the assignments can account for some experimental data to some extent. However, no definite conclusion can be obtained up to now, more theoretical and experimental works are still needed.

In the theoretical scheme of the QCD sum rules having  our unique feature,
we   have performed comprehensive analysis of the exotic states \cite{WZG-review}.  For example, the hidden-charm tetraquark states with the  $J^{PC}=0^{++}$,  $0^{-\pm}$, $1^{-\mp}$,  $1^{+\mp}$,  $2^{++}$ \cite{WZG-HC-spectrum-NPB,WZG-tetra-psedo-NPB,WZG-HC-spectrum-PRD,
WZG-EPJC-P-wave,WZG-EPJC-P-2P}, hidden-bottom tetraquark states with the  $J^{PC}=0^{++}$, $1^{+\mp}$,  $2^{++}$ \cite{WZG-HB-spectrum-EPJC}, hidden-charm molecular states with the  $J^{PC}=0^{++}$, $1^{+\mp}$,  $2^{++}$ \cite{WZG-mole-IJMPA}, doubly-charm tetraquark (molecular) states  with the  $J^{P}=0/1/2^{+}$ \cite{WZG-tetra-cc-EPJC} (\cite{WZG-XQ-mole-EPJA}), hidden-charm pentaquark (molecular) states with the $J^{P}={\frac{1}{2}}/{\frac{3}{2}}/{\frac{5}{2}}^{-}$ \cite{WZG-penta-cc-IJMPA-2050003}(\cite{XWWang-penta-mole}).

In Ref.\cite{WZG-HC-spectrum-NPB}, we take account of the light-flavor  $SU(3)$ breaking effects comprehensively, and revisit the assignments of the potential  $cs\bar{c}\bar{s}$ tetraquark candidates and supersede the old assignments \cite{WangZG-Di-Y4140,WZG-X4140-X4685,X3915-X4500-EPJA-WZG,X3915-X4500-EPJC-WZG}. The possible assignments of the $X(4140)$, $X(4274)$, $X(4500)$, $X(4685)$ and $X(4700)$ in terms of the $cs\bar{c}\bar{s}$ tetraquark states
are given in Table \ref{Identifications-Table-cscs-positive}.  Compared with  the $cu\bar{c}\bar{d}$ or $cu\bar{c}\bar{u}-cd\bar{c}\bar{d}$ tetraquark states \cite{WZG-HC-spectrum-PRD}, the light-flavor $SU(3)$ breaking effects on the tetraquark masses are remarkable.
If the $X(4500)$, $X(4685)$, $X(4700)$ and $X(4630)$ have the same Dirac spinor  structures as the $X(4475)$, $X(4650)$, $X(4710)$ and $X(4800)$ respectively, but different isospin structures, the light-flavor $SU(3)$ breaking effects on the hadron masses should be very small, we would like to explore the odd phenomenon.

\begin{table}
\begin{center}
\begin{tabular}{|c|c|c|c|c|c|c|c|c|}\hline\hline
                              &$J^{PC}$     & $X(1\rm{S})$    &$X(2\rm{S})$ \\ \hline

$[sc]_{A}[\overline{sc}]_{A}$  &$0^{++}$     &       &?\,$X(4700)$ \\ \hline

$[sc]_{\tilde{A}}[\overline{sc}]_{\tilde{A}}$  &$0^{++}$    &                     &?\,$X(4700)$ \\ \hline

$[sc]_{S}[\overline{sc}]_{S}{}^*$  &$0^{++}$     &?\,$X(3960)$      &?\,$X(4500)$   \\ \hline

$[sc]_{V}[\overline{sc}]_{V}{}^*$  &$0^{++}$    &?\,$X(4700)$  & \\ \hline

$[sc]_S[\overline{sc}]_{A}+[sc]_{A}[\overline{sc}]_S$   &$1^{++}$       &?\,$X(4140)$       &?\,$X(4685)$   \\ \hline

$[sc]_S[\overline{sc}]_{\widetilde{A}}+[sc]_{\widetilde{A}}[\overline{sc}]_S$
 &$1^{++}$       &?\,$X(4140)$  &?\,$X(4685)$   \\ \hline

$[sc]_{\widetilde{V}}[\overline{sc}]_{V}-[sc]_{V}[\overline{sc}]_{\widetilde{V}}$
&$1^{++}$       &?\,$X(4274)$ &    \\
\hline\hline
\end{tabular}
\end{center}
\caption{ The possible  assignments of the $cs\bar{c}\bar{s}$ tetraquark states \cite{WZG-HC-spectrum-NPB}, where the subscript $*$ denotes that different energy gaps between the 1S and 2S states are chosen. }\label{Identifications-Table-cscs-positive}
\end{table}

In Ref.\cite{WZG-IJMPA-VV},   we introduce an explicit P-wave to construct the doubly-charm diquarks $\varepsilon^{ijk} c^{T}_j C\gamma_5\stackrel{\leftrightarrow}{\partial}_\mu c_k$, then we take them  as the basic constituents to study the  fully-charm tetraquark states with the QCD sum rules, where   the derivative  $\stackrel{\leftrightarrow}{\partial}_\mu=\stackrel{\rightarrow}{\partial}_\mu-\stackrel{\leftarrow}{\partial}_\mu$ embodies  the net P-wave effect.
 We can take the vector heavy-light  diquarks $\varepsilon^{ijk} c^{T}_j C\gamma_5\stackrel{\leftrightarrow}{D}_\mu q_k$  with $D_\mu=\partial_\mu-ig_sG_\mu$ as the elementary constituents  to study  the  $cq\bar{c}\bar{q}^\prime$ tetraquark states with the  $J^{PC}=0^{++}$, $1^{+-}$ and $2^{++}$,  examine the
light-flavor $SU(3)$ breaking effects, and make reasonable  assignments of the  new $X$ states. In the heavy quark limit, the $c$-quark is static, the  diquarks $\varepsilon^{ijk} c^{T}_j C\gamma_5\stackrel{\leftrightarrow}{D}_\mu q_k$ are reduced to the form $\varepsilon^{ijk} c^{T}_j C\gamma_5D_\mu q_k$, and we would like to take the reduced operators and denote them as $\widehat{V}$.

In Ref.\cite{X4140-X4700-tetra-ChenHX}, Chen et al choose a D-wave diquark  in the color $\bar{\mathbf{3}}$ ($\mathbf{6}$)  and a S-wave antidiquark  in the color $\mathbf{3}$ ($\bar{\mathbf{6}}$) to construct the four-quark current to study the tetraquark state with the $J^P=0^+$, and obtain the tetraquark mass $4.55^{+0.19}_{-0.13}\,\rm{GeV}$ ($4.66^{+0.20}_{-0.14}\,\rm{GeV}$), then assign  the $X(4500)$ and $X(4700)$ tentatively. While in this work, we choose the $[cq]_{\widehat{V}}[\bar{c}\bar{q}^\prime]_{\widehat{V}}$-type currents, which have two P-waves (an effective D-wave) instead of a D-wave.

The article is arranged as follows:  we obtain the QCD sum rules for the  hidden-charm tetraquark states in section 2; in section 3, we   present the numerical results and discussions; section 4 is reserved for our conclusion.

\section{QCD sum rules for  the  hidden-charm  tetraquark states}
At the beginning, we write down  the two-point correlation functions $\Pi(p)$ and $\Pi_{\mu\nu\alpha\beta}(p)$,
\begin{eqnarray}\label{CF-Pi}
\Pi(p)&=&i\int d^4x e^{ip \cdot x} \langle0|T\Big\{J(x)J^{\dagger}(0)\Big\}|0\rangle \, ,\nonumber\\
\Pi_{\mu\nu\alpha\beta}(p)&=&i\int d^4x e^{ip \cdot x} \langle0|T\Big\{J_{\mu\nu}(x)J_{\alpha\beta}^{\dagger}(0)\Big\}|0\rangle \, ,
\end{eqnarray}
where $J_{\mu\nu}(x)=J^{\pm}_{\mu\nu}(x)$,
\begin{eqnarray}
J(x)&=&\varepsilon^{ijk}\varepsilon^{imn}D_\mu q^{T}_j(x)C\gamma_5  c_k(x) \, D^\mu\bar{q}^\prime_m(x) \gamma_5 C \bar{c}^{T}_n(x) \, ,
\end{eqnarray}
\begin{eqnarray}
J^{\pm}_{\mu\nu}(x)&=&\frac{\varepsilon^{ijk}\varepsilon^{imn}}{\sqrt{2}}
\Big[D_\mu q^{T}_j(x)C\gamma_5 c_k(x)\, D_\nu\bar{q}^\prime_m(x)\gamma_5 C \bar{c}^{T}_n(x)\pm \nonumber\\
 &&D_\nu q^{T}_j(x)C\gamma_5  c_k(x)D_\mu\bar{q}^\prime_m(x) \gamma_5C \bar{c}^{T}_n(x) \Big] \, ,
\end{eqnarray}
 the $i$, $j$, $k$, $m$, $n$ are  color indexes, $q$, $q^\prime=u$, $d$ or $s$,
 the  charge conjugation matrix $C=i \gamma^2 \gamma^0$, the superscripts $\pm$ denote  the $\pm$ charge conjugation, respectively.
We  check the properties under parity $\widehat{P}$ and charge-conjugation $\widehat{C}$ transformations,
\begin{eqnarray}\label{P-trans}
\widehat{P} J(x)\widehat{P}^{-1}&=&+J(\tilde{x}) \, , \nonumber\\
\widehat{P} J^{\pm}_{\mu\nu}(x)\widehat{P}^{-1}&=&+J_{\pm}^{\mu\nu}(\tilde{x}) \, ,
\end{eqnarray}
 and
\begin{eqnarray}\label{C-trans}
\widehat{C}J(x)\widehat{C}^{-1}&=&+ J(x) \, , \nonumber\\
\widehat{C}J^{\pm}_{\mu\nu}(x)\widehat{C}^{-1}&=&\pm J^{\pm}_{\mu\nu}(x)  \, ,
\end{eqnarray}
where the coordinates $x^\mu=(t,\vec{x})$ and $\tilde{x}^\mu=(t,-\vec{x})$.

In the isospin limit, the currents with the  symbolic quark constituents $\bar{c}c\bar{d}u$, $\bar{c}c\bar{u}d$, $\bar{c}c\frac{\bar{u}u-\bar{d}d}{\sqrt{2}}$, $\bar{c}c\frac{\bar{u}u+\bar{d}d}{\sqrt{2}}$ couple potentially  to the tetraquark states with degenerated  masses. The currents with the isospins  $I=1$ and $0$ lead to the same QCD sum rules, while the currents with the  symbolic quark constituents $\bar{c}c\bar{q}s$ and $\bar{c}c\bar{s}q$ (with $q=u$ and $d$) couple potentially  to the tetraquark states with degenerated  masses, and lead to the same QCD sum rules. In this work, we would like to choose the quark configurations $\bar{c}c\bar{d}u$, $\bar{c}c\bar{s}q$ and $\bar{c}c\bar{s}s$ to explore the mass spectrum.

Now we insert  a complete set of intermediate hadronic states with the same quantum numbers as the  currents  into the correlation functions   to obtain  the hadronic representation, then isolate the ground state contributions,
\begin{eqnarray}
\Pi(p)&=&\frac{\lambda_{X^+}^2}{M_{X^+}^2-p^2} +\cdots =\Pi_{+}(p^2) \, ,\nonumber\\
\end{eqnarray}
\begin{eqnarray}
\Pi^{-}_{\mu\nu\alpha\beta}(p)&=&\frac{\lambda_{ X^+}^2}{M_{X^+}^2\left(M_{X^+}^2-p^2\right)}\left(p^2g_{\mu\alpha}g_{\nu\beta} -p^2g_{\mu\beta}g_{\nu\alpha} -g_{\mu\alpha}p_{\nu}p_{\beta}-g_{\nu\beta}p_{\mu}p_{\alpha}+g_{\mu\beta}p_{\nu}p_{\alpha}+g_{\nu\alpha}p_{\mu}p_{\beta}\right) \nonumber\\
&&+\frac{\lambda_{ X^-}^2}{M_{X^-}^2\left(M_{X^-}^2-p^2\right)}\left( -g_{\mu\alpha}p_{\nu}p_{\beta}-g_{\nu\beta}p_{\mu}p_{\alpha}+g_{\mu\beta}p_{\nu}p_{\alpha}+g_{\nu\alpha}p_{\mu}p_{\beta}\right) +\cdots  \nonumber\\
&=&\widetilde{\Pi}_{+}(p^2)\left(p^2g_{\mu\alpha}g_{\nu\beta} -p^2g_{\mu\beta}g_{\nu\alpha} -g_{\mu\alpha}p_{\nu}p_{\beta}-g_{\nu\beta}p_{\mu}p_{\alpha}+g_{\mu\beta}p_{\nu}p_{\alpha}+g_{\nu\alpha}p_{\mu}p_{\beta}\right) \nonumber\\
&&+\widetilde{\Pi}_{-}(p^2)\left( -g_{\mu\alpha}p_{\nu}p_{\beta}-g_{\nu\beta}p_{\mu}p_{\alpha}+g_{\mu\beta}p_{\nu}p_{\alpha}+g_{\nu\alpha}p_{\mu}p_{\beta}\right) \, ,\nonumber\\
\Pi_{\mu\nu\alpha\beta}^{+}(p)&=&\frac{\lambda_{ X^+}^2}{M_{X^+}^2-p^2}\left( \frac{\widetilde{g}_{\mu\alpha}\widetilde{g}_{\nu\beta}+\widetilde{g}_{\mu\beta}\widetilde{g}_{\nu\alpha}}{2}-\frac{\widetilde{g}_{\mu\nu}\widetilde{g}_{\alpha\beta}}{3}\right) +\cdots \, \, , \nonumber \\
&=&\Pi_{+}(p^2)\left( \frac{\widetilde{g}_{\mu\alpha}\widetilde{g}_{\nu\beta}+\widetilde{g}_{\mu\beta}\widetilde{g}_{\nu\alpha}}{2}-\frac{\widetilde{g}_{\mu\nu}\widetilde{g}_{\alpha\beta}}{3}\right) +\cdots\, ,
\end{eqnarray}
where $\widetilde{g}_{\mu\nu}=g_{\mu\nu}-\frac{p_{\mu}p_{\nu}}{p^2}$.  We add the superscripts $\pm$ in the $\Pi^{\pm}_{\mu\nu\alpha\beta}(p)$ to denote  $\pm$ charge conjugation, respectively,  and  add the superscripts/subscripts $\pm$ in the   $X^{\pm}$/$\Pi_{\pm}(p^2)$/$\widetilde{\Pi}_{\pm}(p^2)$ to denote  $\pm$ parity, respectively.  The pole residues  $\lambda_{X^\pm}$ are defined by
\begin{eqnarray}
 \langle 0|J(0)|X^+(p)\rangle &=&\lambda_{X^+}\, , \nonumber\\
 \langle 0|J_{\mu\nu}^{+}(0)|X^+(p)\rangle &=& \frac{\lambda_{X^+}}{M_{X^+}} \, \varepsilon_{\mu\nu\alpha\beta} \, \varepsilon^{\alpha}p^{\beta}\, , \nonumber\\
  \langle 0|J^{-}_{\mu\nu}(0)|X^-(p)\rangle &=&\frac{\lambda_{X^-}}{M_{X^-}} \left(\varepsilon_{\mu}p_{\nu}-\varepsilon_{\nu}p_{\mu} \right)\, , \nonumber\\
  \langle 0|J_{\mu\nu}^{+}(0)|X^+(p)\rangle &=& \lambda_{X^+}\, \varepsilon_{\mu\nu} \, ,
\end{eqnarray}
where the  $\varepsilon_{\mu/\alpha}$ and $\varepsilon_{\mu\nu}$ are the tetraquark  polarization vectors.

We choose the components $\Pi_{+}(p^2)$ and $p^2\widetilde{\Pi}_{+}(p^2)$ to explore the  tetraquark states with the $J^{PC}=0^{++}$, $1^{+-}$ and $2^{++}$ without contaminations from  other states. Generally speaking, the quantum field theory does not forbid the currents $J(x)$ and $J_{\mu\nu}(x)$ coupling  to the two-meson scattering states if they have the same quantum numbers, there might  exist contaminations from the two-meson scattering states.
In Refs.\cite{WangZG-Landau,Wang-Two-particle}, we  illustrate that  the two-meson scattering states play an unimportant role and cannot saturate the QCD sum rules by themselves,  on the other hand,  the tetraquark (molecular) states play an irreplaceable role. We can saturate the QCD sum rules with or without the two-particle scattering states, it is reliable to study the  tetraquark (molecular) states. In other words, we choose the local four-quark currents, and the mesons have finite average radii $\sqrt{\langle r^2\rangle}$, the net effects of the small overlapping  of the wave-functions can be absorbed into  the pole residues safely \cite{WangZG-Local}.

At the QCD side, we accomplish the operator product expansion  up to the condensates of dimension $10$ consistently based on our unique counting rules and adopt the truncation ${\mathcal{O}}(\alpha_s^k)$ with $k\leq 1$ for the quark-gluon operators so as to estimate  reliability and feasibility \cite{WZG-tetra-psedo-NPB,WZG-HC-spectrum-PRD,WZG-EPJC-P-wave,WZG-EPJC-P-2P}. We compute the condensates $\langle\bar{q}q\rangle$, $\langle\frac{\alpha_{s}GG}{\pi}\rangle$, $\langle\bar{q}g_{s}\sigma Gq\rangle$, $\langle\bar{q}q\rangle^2$,
$\langle\bar{q}q\rangle \langle\frac{\alpha_{s}GG}{\pi}\rangle$,  $\langle\bar{q}q\rangle  \langle\bar{q}g_{s}\sigma Gq\rangle$,
$\langle\bar{q}g_{s}\sigma Gq\rangle^2$ and $\langle\bar{q}q\rangle^2 \langle\frac{\alpha_{s}GG}{\pi}\rangle$ with $q=u$, $d$ or $s$. Then  we obtain the QCD spectral densities $\rho_{QCD}(s)$ through dispersion relation directly.
In computations, we take the full quark propagators,
\begin{eqnarray}
S^{ij}(x)&=& \frac{i\delta_{ij}\!\not\!{x}}{ 2\pi^2x^4}
-\frac{\delta_{ij}m_q}{4\pi^2x^2}-\frac{\delta_{ij}\langle
\bar{q}q\rangle}{12} +\frac{i\delta_{ij}\!\not\!{x}m_q
\langle\bar{q}q\rangle}{48}-\frac{\delta_{ij}x^2\langle \bar{q}g_s\sigma Gq\rangle}{192}+\frac{i\delta_{ij}x^2\!\not\!{x} m_q\langle \bar{q}g_s\sigma
 Gq\rangle }{1152}\nonumber\\
&& -\frac{ig_s G^{a}_{\alpha\beta}t^a_{ij}(\!\not\!{x}
\sigma^{\alpha\beta}+\sigma^{\alpha\beta} \!\not\!{x})}{32\pi^2x^2} -\frac{\delta_{ij}x^4\langle \bar{q}q \rangle\langle g_s^2 GG\rangle}{27648}-\frac{1}{8}\langle\bar{q}_j\sigma^{\mu\nu}q_i \rangle \sigma_{\mu\nu}  +\cdots \, ,
\end{eqnarray}
\begin{eqnarray}
S_Q^{ij}(x)&=&\frac{i}{(2\pi)^4}\int d^4k e^{-ik \cdot x} \left\{
\frac{\delta_{ij}}{\!\not\!{k}-m_Q}
-\frac{g_sG^n_{\alpha\beta}t^n_{ij}}{4}\frac{\sigma^{\alpha\beta}(\!\not\!{k}+m_Q)+(\!\not\!{k}+m_Q)
\sigma^{\alpha\beta}}{(k^2-m_Q^2)^2}\right.\nonumber\\
&&\left. +\frac{g_s D_\alpha G^n_{\beta\lambda}t^n_{ij}(f^{\lambda\beta\alpha}+f^{\lambda\alpha\beta}) }{3(k^2-m_Q^2)^4}-\frac{g_s^2 (t^at^b)_{ij} G^a_{\alpha\beta}G^b_{\mu\nu}(f^{\alpha\beta\mu\nu}+f^{\alpha\mu\beta\nu}+f^{\alpha\mu\nu\beta}) }{4(k^2-m_Q^2)^5}+\cdots\right\} \, ,\nonumber\\
f^{\lambda\alpha\beta}&=&(\!\not\!{k}+m_Q)\gamma^\lambda(\!\not\!{k}+m_Q)\gamma^\alpha(\!\not\!{k}+m_Q)\gamma^\beta(\!\not\!{k}+m_Q)\, ,\nonumber\\
f^{\alpha\beta\mu\nu}&=&(\!\not\!{k}+m_Q)\gamma^\alpha(\!\not\!{k}+m_Q)\gamma^\beta(\!\not\!{k}+m_Q)\gamma^\mu(\!\not\!{k}+m_Q)\gamma^\nu(\!\not\!{k}+m_Q)\, ,
\end{eqnarray}
with $Q=c$, $t^n=\frac{\lambda^n}{2}$, the $\lambda^n$ is the Gell-Mann matrix \cite{WangHuangTao-3900,Reinders85,Pascual-1984}.
We introduce the new terms $\langle\bar{q}_j\sigma_{\mu\nu}q_i \rangle$  in the full light-quark propagators through  the Fierz re-arrangement  \cite{WangHuangTao-3900}, which  absorb the gluons  emitted from the other  quark lines and result in the mixed condensate $\langle\bar{q}g_s\sigma G q\rangle$.

We match  the hadron side with the QCD  side for the components $\Pi_{+}(p^2)$ and $p^2\widetilde{\Pi}_{+}(p^2)$ in the spectral representation below the continuum thresholds   $s_0$ and carry out the Borel transformation   in regard to
the $P^2=-p^2$ to obtain   the  QCD sum rules:
\begin{eqnarray}\label{QCDSR}
\lambda^2_{X^+}\, \exp\left(-\frac{M^2_{X^+}}{T^2}\right)= \int_{4m_c^2}^{s_0} ds\, \rho_{QCD}(s) \, \exp\left(-\frac{s}{T^2}\right) \, ,
\end{eqnarray}
where the $T^2$ is the Borel parameter.

As last, we differentiate the  QCD sum rules in Eq.\eqref{QCDSR} with respect to  the variable $\tau=\frac{1}{T^2}$,  and obtain the QCD sum rules for  the masses of the $cq\overline{cq}$  tetraquark states $X^+$ with the positive parity,
 \begin{eqnarray}
 M^2_{X^+}&=& -\frac{\int_{4m_c^2}^{s_0} ds\frac{d}{d \tau}\rho_{QCD}(s)\exp\left(-\tau s \right)}{\int_{4m_c^2}^{s_0} ds \rho_{QCD}(s)\exp\left(-\tau s\right)}\, .
\end{eqnarray}

\section{Numerical results and discussions}
At first, we write down the energy-scale dependence of  the input parameters from the re-normalization group equation with the lowest order approximation,
\begin{eqnarray}
\langle\bar{q}q \rangle(\mu)&=&\langle\bar{q}q \rangle({\rm 1GeV})\left[\frac{\alpha_{s}({\rm 1GeV})}{\alpha_{s}(\mu)}\right]^{\frac{12}{33-2n_f}}\, , \nonumber\\
 \langle\bar{q}g_s \sigma Gq \rangle(\mu)&=&\langle\bar{q}g_s \sigma Gq \rangle({\rm 1GeV})\left[\frac{\alpha_{s}({\rm 1GeV})}{\alpha_{s}(\mu)}\right]^{\frac{2}{33-2n_f}}\, , \nonumber\\
 m_c(\mu)&=&m_c(m_c)\left[\frac{\alpha_{s}(\mu)}{\alpha_{s}(m_c)}\right]^{\frac{12}{33-2n_f}} \, ,\nonumber\\
m_q(\mu)&=&m_q({\rm 2GeV} )\left[\frac{\alpha_{s}(\mu)}{\alpha_{s}({\rm 2GeV})}\right]^{\frac{12}{33-2n_f}}\, ,
\end{eqnarray}
 where the quarks $q=u$, $d$ and $s$ \cite{PDG,Narison-mix}, the strong fine-structure constant $\alpha_s(\mu)$ is determined in most cases in the next-to-next-to-leading order approximation,
\begin{eqnarray}
\alpha_s(\mu)&=&\frac{1}{b_0t}\left[1-\frac{b_1}{b_0^2}\frac{\log t}{t} +\frac{b_1^2(\log^2{t}-\log{t}-1)+b_0b_2}{b_0^4t^2}\right]\, ,
\end{eqnarray}
with   $t=\log \frac{\mu^2}{\Lambda_{QCD}^2}$, $b_0=\frac{33-2n_f}{12\pi}$, $b_1=\frac{153-19n_f}{24\pi^2}$, $b_2=\frac{2857-\frac{5033}{9}n_f+\frac{325}{27}n_f^2}{128\pi^3}$,  $\Lambda_{QCD}=210\,\rm{MeV}$, $292\,\rm{MeV}$  and  $332\,\rm{MeV}$ for the flavors  $n_f=5$, $4$ and $3$, respectively  \cite{PDG}. And we choose $n_f=4$ in the present analysis.

 At the initial  points, we adopt  the standard values  $\langle
\bar{q}q \rangle=-(0.24\pm 0.01\, \rm{GeV})^3$, $\langle
\bar{s} s \rangle=(0.8 \pm 0.1)\langle \bar{q}q \rangle$,
$\langle \bar{q}g_s\sigma G q \rangle=m_0^2\langle \bar{q}q \rangle$,
$\langle \bar{s}g_s\sigma G s \rangle=m_0^2\langle \bar{s}s \rangle$,
$m_0^2=(0.8 \pm 0.1)\,\rm{GeV}^2$,  $\langle \frac{\alpha_s
GG}{\pi}\rangle=(0.012\pm0.004)\,\rm{GeV}^4 $    at the particular energy scale  $\mu=1\, \rm{GeV}$ with $q=u$ and $d$
\cite{Reinders85,SVZ79,Colangelo-Review}, and adopt  the
modified-minimal-subtraction masses   $m_{c}(m_c)=(1.275\pm0.025)\,\rm{GeV}$ and $m_{s}({\rm 2 GeV})=(0.095\pm0.005)\,\rm{GeV}$ from the Particle Data Group \cite{PDG}. Moreover, we set $m_u=m_d=0$ considering their  tiny values.

In our previous works, we adopt the modified energy scale formula $\mu=\sqrt{M^2_{X/Y/Z}-(2{\mathbb{M}}_c)^2}-\kappa\,{\mathbb{M}}_s$ to choose the suitable   energy scales of the QCD spectral densities \cite{WZG-HC-spectrum-NPB,WZG-NPB-cscs,Wang-tetra-formula,WZG-mole-IJMPA,WZG-XQ-mole-penta,WZG-XQ-mole-EPJA,
WZG-tetra-psedo-NPB,WZG-Zcs3985-4110},
 where the ${\mathbb{M}}_c$ and ${\mathbb{M}}_s$ are the effective $c$ and $s$-quark masses respectively, and have common values. We take ${\mathbb{M}}_c=1.82\,\rm{GeV}$ \cite{WangEPJC-1601} and  ${\mathbb{M}}_s=0.20\,\rm{GeV}$ ($0.12\,\rm{GeV}$) for the S-wave (P-wave) tetraquark  states  \cite{WZG-HC-spectrum-NPB,WZG-NPB-cscs,WZG-mole-IJMPA,WZG-XQ-mole-penta,WZG-XQ-mole-EPJA,
WZG-tetra-psedo-NPB,WZG-Zcs3985-4110}.

In the picture  of  tetraquark states, we can assign  the
($X(3960)$, $X(4500)$), ($X(4140)$,  $X(4685)$), ($Z_c(3900)$, $Z_c(4430)$) and
  ($Z_c(4020)$, $Z_c(4600)$)  as  the (1S, 2S)   tetraquark states tentatively, see Table \ref{Identifications-1S-2S}. We draw the conclusion tentatively  that the mass gaps between the 1S and 2S tetraquark states  are about $0.55\sim 0.59 \,\rm{GeV}$.

\begin{table}
\begin{center}
\begin{tabular}{|c|c|c|c|c|c|c|c|c|}\hline\hline
                              &$J^{PC}$     & $X(1\rm{S})$    &$X(2\rm{S})$ & References \\ \hline

$[sc]_{S}[\overline{sc}]_{S}{}^*$  &$0^{++}$     &?\,$X(3960)$      &?\,$X(4500)$   & \cite{WZG-HC-spectrum-NPB}\\ \hline

$[sc]_S[\overline{sc}]_{A}+[sc]_{A}[\overline{sc}]_S$   &$1^{++}$       &?\,$X(4140)$       &?\,$X(4685)$  &\cite{WZG-HC-spectrum-NPB}  \\ \hline

$[sc]_S[\overline{sc}]_{\widetilde{A}}+[sc]_{\widetilde{A}}[\overline{sc}]_S$
 &$1^{++}$       &?\,$X(4140)$  &?\,$X(4685)$  &\cite{WZG-HC-spectrum-NPB}  \\ \hline

$[uc]_S[\bar{d}\bar{c}]_A- [uc]_A[\bar{d}\bar{c}]_S$
 &$1^{+-}$       &?\,$Z_c(3900)$  &?\,$Z_c(4430)$  &\cite{WangZG-Z4430-CTP,Maiani-II-type,Nielsen-1401}  \\ \hline

 $[uc]_{A}[\overline{dc}]_{A}$ &$1^{+-}$       &?\,$Z_c(4020)$  &?\,$Z_c(4600)$  &\cite{ChenHX-Z4600-A,WangZG-axial-Z4600}  \\ \hline

 $[uc]_S[\overline{dc}]_{\widetilde{A}}-[uc]_{\widetilde{A}}[\overline{dc}]_S$ &$1^{+-}$       &?\,$Z_c(4020)$  &?\,$Z_c(4600)$  &\cite{ChenHX-Z4600-A,WangZG-axial-Z4600}  \\ \hline

$[uc]_{\widetilde{A}}[\overline{dc}]_{A}-[uc]_{A}[\overline{dc}]_{\widetilde{A}}$ &$1^{+-}$       &?\,$Z_c(4020)$  &?\,$Z_c(4600)$  &\cite{ChenHX-Z4600-A,WangZG-axial-Z4600}  \\
\hline\hline
\end{tabular}
\end{center}
\caption{ The possible  assignments of the 1S and 2S tetraquark states. }\label{Identifications-1S-2S}
\end{table}

In the present analysis, we would like to set the continuum threshold parameters as  $\sqrt{s_0}=M_X+0.60\pm 0.10\,\rm{GeV}$, then vary  the continuum threshold parameters  and Borel parameters  to meet with the  four   criteria:\\
$\bullet$ Pole  dominance at the hadron  side;\\
$\bullet$ Convergence of the operator product expansion;\\
$\bullet$ Appearance of the enough flat Borel platforms;\\
$\bullet$ Fulfillment of the  modified energy scale formula,\\
  via trial  and error.
We define the pole contributions (PC),
\begin{eqnarray}
{\rm{PC}}&=&\frac{\int_{4m_{c}^{2}}^{s_{0}}ds\rho_{QCD}\left(s\right)\exp\left(-\frac{s}{T^{2}}\right)} {\int_{4m_{c}^{2}}^{\infty}ds\rho_{QCD}\left(s\right)\exp\left(-\frac{s}{T^{2}}\right)}\, ,
\end{eqnarray}
 and  the contributions of the vacuum condensates $D(n)$ of dimension $n$,
\begin{eqnarray}
D(n)&=&\frac{\int_{4m_{c}^{2}}^{s_{0}}ds\rho_{QCD,n}(s)\exp\left(-\frac{s}{T^{2}}\right)}
{\int_{4m_{c}^{2}}^{s_{0}}ds\rho_{QCD}\left(s\right)\exp\left(-\frac{s}{T^{2}}\right)}\, ,
\end{eqnarray}
in the same way as in our previous works.

Now we add the subscripts $0$, $1$ and $2$ to represent the spins of the tetraquark states. In the light-flavor $SU(3)$ symmetry limit, the QCD spectral densities,
\begin{eqnarray}\label{rho-No-simplify}
\rho_{QCD}^0(s)&\propto& {\rm Pert}\, , \, m_c^2\langle \frac{\alpha_sGG}{\pi}\rangle\, , \, m_q m_c\langle \frac{\alpha_sGG}{\pi}\rangle\, , \, m_q\langle \bar{q}g_s \sigma G q\rangle \, , \, m_c\langle \bar{q}q\rangle\langle \frac{\alpha_sGG}{\pi}\rangle\, , \, m_q\langle \bar{q}q\rangle\langle \frac{\alpha_sGG}{\pi}\rangle\, , \nonumber\\
 &&m_c^2\langle \bar{q}g_s \sigma G q\rangle^2\, , \, m_q m_c\langle \bar{q}g_s \sigma G q\rangle^2\, , \nonumber\\
\rho_{QCD}^{1/2}(s)&\propto& {\rm Pert}\, , \, m_c^2\langle \frac{\alpha_sGG}{\pi}\rangle\, , \, m_q m_c\langle \frac{\alpha_sGG}{\pi}\rangle\, , \, m_c\langle \bar{q}g_s \sigma G q\rangle\, , \, m_q\langle \bar{q}g_s \sigma G q\rangle \, , \, m_c\langle \bar{q}q\rangle\langle \frac{\alpha_sGG}{\pi}\rangle\, , \nonumber\\
  && m_q\langle \bar{q}q\rangle\langle \frac{\alpha_sGG}{\pi}\rangle\, , \,
 m_c^2\langle \bar{q}g_s \sigma G q\rangle^2\, , \, m_q m_c\langle \bar{q}g_s \sigma G q\rangle^2\, ,
\end{eqnarray}
where $q=u$, $d$ or $s$, the Pert denotes the perturbative terms.
All the  condensates are not companied with inverse powers of the Borel parameter $\frac{1}{T^2}$, $\frac{1}{T^4}$  $\cdots$, thus they cannot manifest themselves at small values of $T^2$ to result in very flat platforms. In the QCD sum rules for the $Q\bar{Q}q\bar{q}$ tetraquark states, the gluon condensate always play an  unimportant role \cite{WZG-review}. Furthermore, the vacuum condensates companied with  the light quark mass $m_q$ play a tiny role due to its small value. We can estimate the convergent behavior of the operator product expansion by considering the important terms,
\begin{eqnarray}\label{rho-simplify}
\rho_{QCD}^0(s)&\propto& {\rm Pert}\, , \, m_c^2\langle \bar{q}g_s \sigma G q\rangle^2\, , \nonumber\\
\rho_{QCD}^{1/2}(s)&\propto& {\rm Pert}\, , \, m_c\langle \bar{q}g_s \sigma G q\rangle\, , \,  m_c^2\langle \bar{q}g_s \sigma G q\rangle^2\, .
\end{eqnarray}
 If we require  the  contributions $|D(10)|$ to be about $1\%$,  the dominant contributions would come  from the perturbative terms plus $m_c\langle \bar{q}g_s \sigma G q\rangle$, the condensates $m_c^2\langle \bar{q}g_s \sigma G q\rangle^2$ with $q=u$, $d$ or $s$ almost make no difference.  As there  exists an additional term $m_c\langle \bar{q}g_s \sigma G q\rangle$ in the spectral densities $\rho_{QCD}^{1/2}(s)$, the light-flavor $SU(3)$ breaking effects on the  masses of the tetraquark states with the $J^{PC}=1^{+-}$ and $2^{++}$  would be slightly larger.

In the QCD sum rules from the four-quark currents without explicit P-waves
  \cite{WZG-HC-spectrum-NPB,
  WZG-mole-IJMPA,WZG-XQ-mole-EPJA,WZG-tetra-psedo-NPB,WZG-HC-spectrum-PRD},
  there exist the terms $m_c\langle \bar{q}q\rangle$, $m_c\langle \bar{q}g_s \sigma G q\rangle$, $m_c^2\langle \bar{q}q\rangle^2$, $m_c^2\langle \bar{q}q\rangle\langle \bar{q}g_s \sigma G q\rangle$ and $m_c^2\langle \bar{q}g_s \sigma G q\rangle^2$, some are companied with inverse powers of the Borel parameter $\frac{1}{T^2}$, $\frac{1}{T^4}$  $\cdots$,   the light-flavor $SU(3)$ breaking effects on the tetraquark masses  are normal.
In the present case, due to the tiny light-flavor $SU(3)$ breaking effects, the modified energy scale formula $\mu=\sqrt{M^2_{X/Y/Z}-(2{\mathbb{M}}_c)^2}-\kappa\,{\mathbb{M}}_s$ with $\kappa=0$, $1$ and $2$ cannot be satisfied, we have to set the effective $s$-quark mass ${\mathbb{M}}_s$ to be zero, in other words, we have to resort to the energy scale formula $\mu=\sqrt{M^2_{X/Y/Z}-(2{\mathbb{M}}_c)^2}$ to restrict the QCD sum rules.

After tedious  trial and error, at last, we obtain the Borel windows, continuum threshold parameters, energy scales of the spectral densities,  pole contributions, and  contributions of the  condensates of dimension $10$, and  show them explicitly  in Table \ref{BorelP-ss}. From Table \ref{BorelP-ss},  we can see
explicitly  that the  ground state contributions are about $(37-66)\%$ for the  tetraquark states with the $J^{PC}=0^{++}$ and $(33-57)\%$ for the   tetraquark states with the $J^{PC}=1^{+-}$ and $2^{++}$.  For the tetraquark states with the $J^{PC}=0^{++}$, the pole dominance is satisfied  certainly, while
for the  tetraquark states with the $J^{PC}=1^{+-}$ and $2^{++}$, the pole dominance is only satisfied  marginally. As the largest power of the QCD spectral densities $\rho(s)\sim s^6$ (for the tetraquark states with explicit two P-waves or a D-wave) instead of $s^4$ (for the tetraquark states without explicit P/D-waves), the pole dominance criterion is difficult to satisfy. In Ref.\cite{X4140-X4700-tetra-ChenHX}, the pole contributions only $\geq 20\%$ in the QCD sum rules for the $X(4500)$ and $X(4700)$.
On the other hand,
the contributions of the vacuum condensates
 $D(0)\sim 100\%$ for the tetraquark states with the $J^{PC}=0^{++}$ and $D(0)+D(3)\sim 100\%$ for the tetraquark states with the $J^{PC}=1^{+-}$ and $2^{++}$,    the operator product  expansion is convergent in all the QCD sum rules.

We take  account of all uncertainties of the  parameters and obtain  the masses and pole residues of the  hidden-charm tetraquark states with the  $J^{PC}=0^{++}$, $1^{+-}$ and $2^{++}$, and show them  explicitly in Table \ref{mass-Table-ss}.
The uncertainties of the  masses and pole residues $\Delta f$ are approximately estimated via,
\begin{eqnarray}
\left(\Delta f \right)^2 &=&\sum_i\left(\frac{\partial f}{\partial
x_i}\right)^2\mid_{x_i=\bar{x}_i} (x_i-\bar{x}_i)^2\,  , \nonumber\\
&\approx&\sum_i \left[f(\bar x_i\pm \Delta x_i)-f(\bar x_i)\right]^2\, ,
\end{eqnarray}
where the $f(x)$ stands for the analytical expressions, the $\bar x_i$ stands for the central values of the relevant  parameters, the $\Delta x_i$ are their uncertainties, and $|f(\bar x_i+ \Delta x_i)-f(\bar x_i)|\approx |f(\bar x_i- \Delta x_i)-f(\bar x_i)|$ as the large energy scales of the QCD spectral densities weaken the sensitivity to the $x_i$.

We try to choose  uniform continuum threshold parameters and energy scales for the tetraquark states with the $J^{PC}=0^{++}$, $1^{+-}$ and $2^{++}$ respectively to scrutinize  the light-flavor $SU(3)$ breaking effects.
From Table \ref{mass-Table-ss}, we can obtain the central values  $\mu=2.64583\,\rm{GeV}$ for the tetraquark states with the $J^{PC}=0^{++}$, $\mu=2.79616\,\rm{GeV}$, $2.82887\,\rm{GeV}$ and $2.86135\,\rm{GeV}$ for the tetraquark states with the $J^{PC}=1^{+-}$, and $\mu=2.8936\,\rm{GeV}$, $2.92563\,\rm{GeV}$ and $2.95745\,\rm{GeV}$ for the tetraquark states with the $J^{PC}=2^{++}$
from  the  energy scale formula $\mu=\sqrt{M^2_{X/Y/Z}-(2{\mathbb{M}}_c)^2}$,
which are consistent with the values shown in Table \ref{BorelP-ss}.
We combine Table \ref{BorelP-ss} with Table \ref{mass-Table-ss}, and obtain the conclusion tentatively that
the light-flavor $SU(3)$ breaking effects on the tetraquark masses are tiny.

\begin{figure}
 \centering
 \includegraphics[totalheight=5cm,width=6cm]{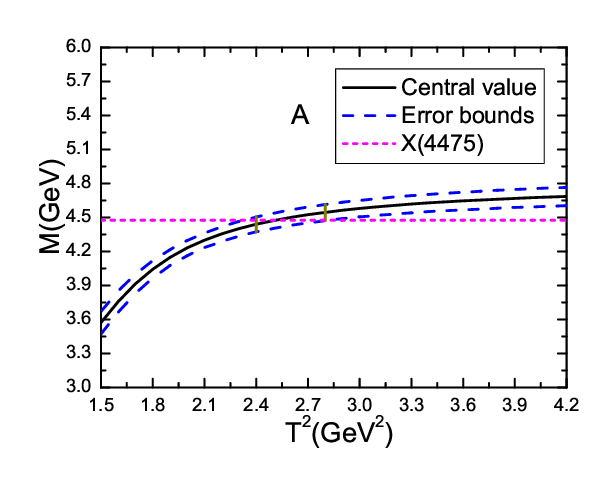}
 \includegraphics[totalheight=5cm,width=6cm]{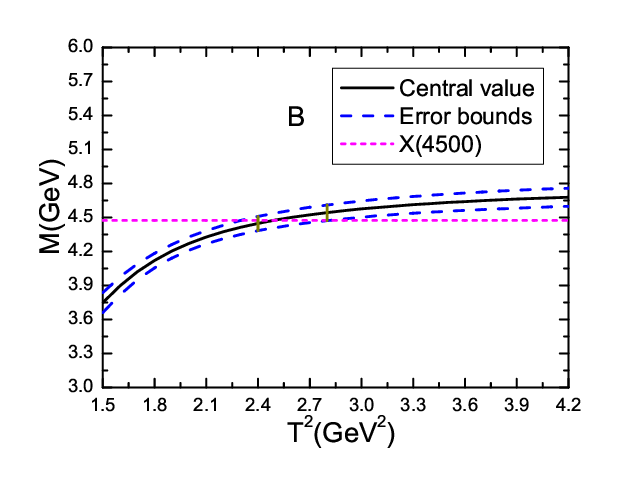}
 \caption{ The masses of the  $[uc]_{\widehat{V}}[\overline{dc}]_{\widehat{V}}$ ($A$) and $[sc]_{\widehat{V}}[\overline{sc}]_{\widehat{V}}$ ($B$) tetraquark states   with the  $J^{PC}=0^{++}$ via variations  of the Borel parameter $T^2$, where the isospin limit is taken.  }\label{mass-1-fig}
\end{figure}

 In  Fig.\ref{mass-1-fig}, we plot the masses of the  $[uc]_{\widehat{V}}[\overline{dc}]_{\widehat{V}}$ and $[sc]_{\widehat{V}}[\overline{sc}]_{\widehat{V}}$ tetraquark states   with the
 $J^{PC}=0^{++}$ according to variations of the Borel parameters at much larger regions than the Borel windows, which are characterized  by  two short vertical  lines.
 For comparison, we present the experimental values of the masses of the $X(4475)$ and $X(4500)$ from the LHCb  collaboration \cite{LHCb-X4685,LHCb-X4500-ccqq}. The predicted masses increase monotonically and quickly  with increase of the Borel parameters,  at the point  $T^2=2.4\,\rm{GeV}^2$, the beginning of the Borel windows, the masses begin to increase slowly, it is feasible to fix the lower bounds of the Borel windows.
 On the other hand, the PC ($D(10)$) decreases monotonically and steadily (quickly) with increase of the Borel parameters, for example, the lowest value $37\%$ ($3\%$) determines the upper (lower) bound of the Borel window for   the $[uc]_{\widehat{V}}[\overline{dc}]_{\widehat{V}}$ tetraquark state, where the predicted mass happens to vary  slowly.
 Other Borel windows shown in Table \ref{BorelP-ss} are fixed in the same way
 by considering the four criteria of the QCD sum rules. The energy gaps $\sqrt{s_0}-M_X\sim 0.55\,\rm{GeV}$ for the tetraquark states with the $J^{PC}=0^{++}$ and $\sim 0.65\,\rm{GeV}$ for the tetraquark states with the $J^{PC}=1^{+-}$ and $2^{++}$, although the value $0.65\,\rm{GeV}$ is somewhat large \cite{WZG-HC-spectrum-NPB,WangZG-Z4430-CTP,ChenHX-Z4600-A,WangZG-axial-Z4600,
 Maiani-II-type,Nielsen-1401}.

In Table \ref{mass-Table-ss}, we present the possible  assignments of the LHCb's  new tetraquark candidates.
The predictions $M_X=4.50\pm0.12\,\rm{GeV}$ and $4.50\pm0.12\,\rm{GeV}$   support
assigning   the  $X(4475)$ and $X(4500)$  as the
$[uc]_{\widehat{V}}[\overline{uc}]_{\widehat{V}}
-[dc]_{\widehat{V}}[\overline{dc}]_{\widehat{V}}$  and
$[sc]_{\widehat{V}}[\overline{sc}]_{\widehat{V}}$ tetraquark states with the $J^{PC}=0^{++}$  respectively.
  The central values of the masses of the
$[uc]_{\widehat{V}}[\overline{dc}]_{\widehat{V}}$,
$[qc]_{\widehat{V}}[\overline{sc}]_{\widehat{V}}$ and $[sc]_{\widehat{V}}[\overline{sc}]_{\widehat{V}}$ tetraquark states with the  $J^{PC}=0^{++}$  are $4.5001\,\rm{GeV}$, $4.5013\,\rm{GeV}$ and $4.5014\,\rm{GeV}$, respectively, the light-flavor $SU(3)$ breaking effects on the tetraquark  masses are extremely tiny.

In Ref.\cite{WZG-HC-spectrum-NPB}, we tentatively assign
the $X(3960)$ and $X(4500)$  as the 1S and 2S $[sc]_{S}[\overline{sc}]_{S}{}^*$  tetraquark states with the  $J^{PC}=0^{++}$  respectively based on the QCD sum rules, where the light-flavor
$SU(3)$ breaking effect   $M_{X(4140)}-M_{X(3872)}=275\,\rm{MeV}$ is consistent with the mass difference $m_s-m_q=135\,\rm{MeV}$ and is normal \cite{PDG}.
So the $X(4500)$ might  have two significant Fock components at least. On the other hand, it is also possible to assign the $X(3960)$ as the $D_s\bar{D}_s$ molecular state based on the QCD sum rules \cite{WZG-XinQ-AAPPS-X3960}.

Also in Ref.\cite{WZG-HC-spectrum-NPB}, we tentatively assign the $X(4700)$  as  the 2S $[sc]_{A}[\overline{sc}]_{A}$/$[sc]_{\tilde{A}}[\overline{sc}]_{\tilde{A}}$  or  1S $[sc]_{V}[\overline{sc}]_{V}{}^*$ tetraquark state with the  $J^{PC}=0^{++}$, see Table \ref{Identifications-Table-cscs-positive}. In case of the constituent diquarks without an explicit P-wave,  the light-flavor $SU(3)$ breaking effects are normal, the tetraquark states with the  symbolic valence structures $c\bar{c}(u\bar{u}-d\bar{d})$ lie about $200\,\rm{MeV}$ below the corresponding $c\bar{c}s\bar{s}$ tetraquark states,  there is no room for the $X(4710)$ \cite{WZG-HC-spectrum-PRD}. In the present case, there is also no room for the $X(4710)$, see Table \ref{mass-Table-ss}. We should resort to other diquarks with an explicit P-wave to construct the four-quark currents to interpolate the $X(4710)$ and $X(4700)$ consistently.

\begin{table}
\begin{center}
\begin{tabular}{|c|c|c|c|c|c|c|c|c|}\hline\hline
 $X$($Z$)           &$J^{PC}$ & $T^2 (\rm{GeV}^2)$ & $\sqrt{s_0}(\rm GeV) $      &$\mu(\rm{GeV})$   &pole         &$|D(10)|$ \\ \hline

$[uc]_{\widehat{V}}[\overline{dc}]_{\widehat{V}}$  &$0^{++}$  &$2.4-2.8$  &$5.05\pm0.10$               &$2.6$   &$(37-65)\%$  &$(1\sim 3)\%$   \\
 \hline

$[qc]_{\widehat{V}}[\overline{sc}]_{\widehat{V}}$  &$0^{++}$  &$2.4-2.8$  &$5.05\pm0.10$               &$2.6$   &$(38-66)\%$  &$(1\sim 2)\%$  \\
 \hline

$[sc]_{\widehat{V}}[\overline{sc}]_{\widehat{V}}$  &$0^{++}$  &$2.4-2.8$  &$5.05\pm0.10$               &$2.6$   &$(37-65)\%$  &$(1\sim 1)\%$   \\
 \hline

$[uc]_{\widehat{V}}[\overline{dc}]_{\widehat{V}}$     &$1^{+-}$  &$3.2-3.6$   &$5.25\pm0.10$               &$2.8$   &$(35-57)\%$    &$(1\sim 1)\%$  \\ \hline

$[qc]_{\widehat{V}}[\overline{sc}]_{\widehat{V}}$     &$1^{+-}$  &$3.2-3.6$   &$5.25\pm0.10$               &$2.8$   &$(35-56)\%$    &$(1\sim 1)\%$ \\ \hline

$[sc]_{\widehat{V}}[\overline{sc}]_{\widehat{V}}$     &$1^{+-}$  &$3.2-3.6$   &$5.25\pm0.10$               &$2.9$   &$(34-56)\%$    &$(0\sim 1)\%$ \\ \hline

$[uc]_{\widehat{V}}[\overline{dc}]_{\widehat{V}}$    &$2^{++}$  &$3.3-3.7$   &$5.30\pm0.10$               &$2.9$   &$(34-55)\%$     &$(1\sim 1)\%$\\ \hline

$[qc]_{\widehat{V}}[\overline{sc}]_{\widehat{V}}$    &$2^{++}$  &$3.3-3.7$   &$5.30\pm0.10$               &$2.9$   &$(33-54)\%$     &$(1\sim 1)\%$ \\ \hline

$[sc]_{\widehat{V}}[\overline{sc}]_{\widehat{V}}$    &$2^{++}$  &$3.3-3.7$   &$5.30\pm0.10$               &$3.0$   &$(33-54)\%$     &$(0\sim 1)\%$ \\
\hline\hline
\end{tabular}
\end{center}
\caption{ The Borel windows, continuum threshold parameters, energy scales,  pole contributions, and contributions of the condensates of dimension $10$  for the hidden-charm tetraquark states. }\label{BorelP-ss}
\end{table}

The predictions $M_Z=4.59\pm0.11\,\rm{GeV}$ and $4.61\pm0.11\,\rm{GeV}$    support assigning  the
$Z_{c}(4600)$ and $Z_{\bar{c}\bar{s}}(4600)$ as the
   $[uc]_{\widehat{V}}[\overline{dc}]_{\widehat{V}}$ and $[qc]_{\widehat{V}}[\overline{sc}]_{\widehat{V}}$ tetraquark states with the
$J^{PC}=1^{+-}$ respectively. Again, the light-flavor $SU(3)$ breaking effects on the tetraquark  masses  are tiny.
In Refs.\cite{ChenHX-Z4600-A,WangZG-axial-Z4600,WZG-HC-spectrum-PRD},
  the $Z_c(4020)$ and $Z_c(4600)$ are assigned as the 1S and 2S $[uc]_{A}[\overline{dc}]_{A}$,
$[uc]_S[\overline{dc}]_{\widetilde{A}}-[uc]_{\widetilde{A}}[\overline{dc}]_S$ or
$[uc]_{\widetilde{A}}[\overline{dc}]_{A}-[uc]_{A}[\overline{dc}]_{\widetilde{A}}$  tetraquark states with the  $J^{PC}=1^{+-}$ respectively, see Table \ref{Identifications-1S-2S}. In  Ref.\cite{WZG-HC-spectrum-NPB},  the $X(4140)$ and $X(4685)$ are assigned as the 1S and 2S $[sc]_S[\overline{sc}]_{A}+[sc]_{A}[\overline{sc}]_S$  or
$[sc]_S[\overline{sc}]_{\widetilde{A}}+[sc]_{\widetilde{A}}[\overline{sc}]_S$  tetraquark states with the $J^{PC}=1^{++}$ respectively, also see Table \ref{Identifications-1S-2S}. In Table \ref{Identifications-1S-2S}, the light-flavor $SU(3)$ breaking effects on the tetraquark masses are normal. On the other hand, if the $Z_{cs}(3985/4000)$
is assigned as the 1S $[sc]_A[\overline{qc}]_{A}$ tetraquark state with the $J^{PC}=1^{+-}$ \cite{WZG-Zcs3985-4110}, then the $Z_{\bar{c}\bar{s}}(4600)$ and $Z_{\bar{c}\bar{s}}(4900)$ can be assigned as the 2S and 3S $[qc]_A[\overline{sc}]_{A}$ tetraquark states with the $J^{PC}=1^{+-}$, respectively, such a possibility  also exists. The $Z_c(4600)$ and $Z_{\bar{c}\bar{s}}(4600)$ might have several significant Fock components.

 We can take the pole residues $\lambda_X$ in Table \ref{mass-Table-ss} as  input parameters to explore  the strong decays of those  tetraquark states with the (light-cone) QCD sum rules, and acquire the partial decay widths and branching  fractions to diagnose the nature of those exotic states \cite{WZG-review}.

\begin{table}
\begin{center}
\begin{tabular}{|c|c|c|c|c|c|c|c|c|}\hline\hline
$X$($Z$)   &$J^{PC}$  &$M_X (\rm{GeV})$   &$\lambda_X (\rm{GeV}^7) $ & Assignments  \\ \hline

$[uc]_{\widehat{V}}[\overline{dc}]_{\widehat{V}}$  &$0^{++}$  &$4.50\pm0.12$  &$(4.96\pm1.39)\times 10^{-2}$ & ?\,$X(4475)$  \\ \hline

$[qc]_{\widehat{V}}[\overline{sc}]_{\widehat{V}}$  &$0^{++}$  &$4.50\pm0.12$  &$(5.18\pm1.41)\times 10^{-2}$ & \\ \hline

$[sc]_{\widehat{V}}[\overline{sc}]_{\widehat{V}}$  &$0^{++}$  &$4.50\pm0.12$  &$(5.38\pm1.42)\times 10^{-2}$ &?\,$X(4500)$ \\ \hline

$[uc]_{\widehat{V}}[\overline{dc}]_{\widehat{V}}$  &$1^{+-}$  &$4.59\pm0.11$  &$(2.15\pm0.44)\times 10^{-2}$ &   ?\,$Z_{c}(4600)$ \\ \hline

$[qc]_{\widehat{V}}[\overline{sc}]_{\widehat{V}}$  &$1^{+-}$  &$4.61\pm0.11$  &$(2.20\pm0.45)\times 10^{-2}$  &?\,$Z_{\bar{c}\bar{s}}(4600)$ \\ \hline

$[sc]_{\widehat{V}}[\overline{sc}]_{\widehat{V}}$  &$1^{+-}$  &$4.63\pm0.11$  &$(2.28\pm0.46)\times 10^{-2}$ & \\ \hline

$[uc]_{\widehat{V}}[\overline{dc}]_{\widehat{V}}$  &$2^{++}$  &$4.65\pm0.11$  &$(3.52\pm0.71)\times 10^{-2}$ & \\ \hline

$[qc]_{\widehat{V}}[\overline{sc}]_{\widehat{V}}$  &$2^{++}$  &$4.67\pm0.11$  &$(3.61\pm0.73)\times 10^{-2}$ & \\ \hline

$[sc]_{\widehat{V}}[\overline{sc}]_{\widehat{V}}$  &$2^{++}$  &$4.69\pm0.11$  &$(3.75\pm0.75)\times 10^{-2}$ & \\
\hline\hline
\end{tabular}
\end{center}
\caption{ The masses and pole residues of the hidden-charm  tetraquark states,  where the isospin limit is taken. }\label{mass-Table-ss}
\end{table}

\section{Conclusion}
In this work,  we introduce an explicit P-wave to construct the diquarks, then construct  the local four-quark currents to explore  the  hidden-charm  tetraquark states with the  $J^{PC}=0^{++}$, $1^{+-}$ and $2^{++}$ in the framework of the  QCD sum rules at length. We carry out the operator product expansion up to the  condensates of dimension $10$  in a consistent way. Direct calculations indicate tiny light-flavor $SU(3)$ breaking effects on the tetraquark masses and  the  modified energy scale formula $\mu=\sqrt{M^2_{X/Y/Z}-(2{\mathbb{M}}_c)^2}-\kappa\,{\mathbb{M}}_s$ can  be satisfied only  by setting ${\mathbb{M}}_s=0$.
The present calculations   support
assigning the  $X(4475)$ and $X(4500)$ as the
$[uc]_{\widehat{V}}[\overline{uc}]_{\widehat{V}}-[dc]_{\widehat{V}}[\overline{dc}]_{\widehat{V}}$  and
$[sc]_{\widehat{V}}[\overline{sc}]_{\widehat{V}}$ tetraquark states with the $J^{PC}=0^{++}$,  respectively,   support assigning the $Z_{c}(4600)$ and $Z_{\bar{c}\bar{s}}(4600)$ as the
   $[uc]_{\widehat{V}}[\overline{dc}]_{\widehat{V}}$ and $[qc]_{\widehat{V}}[\overline{sc}]_{\widehat{V}}$
 tetraquark states with the
$J^{PC}=1^{+-}$, respectively. On the other hand, there is no room for the $X(4710)$ and $X(4700)$.  Considering previous works, we can obtain the conclusion tentatively that the $X(4475)$, $X(4500)$, $Z_{c}(4600)$ and $Z_{\bar{c}\bar{s}}(4600)$  might have other important Fock components besides  the $\widehat{V}\widehat{V}$ type components, which could account for the experimental data from the LHCb collaboration.
Other predictions can be confronted to the experimental data in the future  to examine the exotic states.

\section*{Acknowledgements}
This  work is supported by National Natural Science Foundation, Grant Number  12175068.

\section*{Authors' contributions}

I am the single author.

\section*{Funding}

This research was supported by the National Natural Science Foundation of
China through Grant No.12175068.

\section*{Availability of data and materials}
The data  are available via contacting the corresponding author upon request.

\section*{Competing interests}
I declare that I have no competing interests.


\begin{thebibliography}{99}

\bibitem{LHCb-16061}  R. Aaij et al,  Phys. Rev. Lett. {\bf 118} (2017) 022003.

\bibitem{LHCb-16062}  R. Aaij et al,  Phys. Rev. {\bf D95} (2017) 012002.

\bibitem{LHCb-X4685} R. Aaij et al, Phys. Rev. Lett. {\bf 127} (2021) 082001.

\bibitem{WZG-HC-spectrum-NPB} Z. G. Wang, Nucl. Phys. {\bf B1007} (2024) 116661.

\bibitem{WZG-NPB-cscs}  Z. G. Wang, Nucl. Phys. {\bf B1002} (2024) 116514.

\bibitem{WZG-review} Z. G. Wang, arXiv:2502.11351 [hep-ph].


\bibitem{LHCb-X4500-ccqq} R. Aaij et al,  JHEP {\bf 01} (2025) 054.

\bibitem{LHCb-Z4600}  R. Aaij et al, Phys. Rev. Lett. {\bf 122} (2019)  152002.

\bibitem{WangZG-Z4430-CTP} Z. G. Wang,  Commun. Theor. Phys. {\bf 63} (2015)  325.

\bibitem{ChenHX-Z4600-A} H. X. Chen and W. Chen,  Phys. Rev. {\bf D99} (2019)  074022.

\bibitem{WangZG-axial-Z4600} Z. G. Wang, Chin. Phys. {\bf C44} (2020) 063105.



\bibitem{LHCb-X4500-X4274} R. Aaij et al,  Phys. Rev. Lett. {\bf 134} (2025)  031902.




\bibitem{WangHuangTao-3900} Z. G. Wang and T. Huang,  Phys. Rev. {\bf D89} (2014) 054019.

\bibitem{WangTetraquarkCTP} Z. G. Wang, Commun. Theor. Phys. {\bf 63} (2015) 466.

\bibitem{Wang-tetra-formula} Z. G. Wang, Eur. Phys. J. {\bf C74} (2014)  2874.

\bibitem{Wang-Huang-NPA-2014} Z. G. Wang and T. Huang, Nucl. Phys. {\bf A930} (2014) 63.

\bibitem{WZG-IJMPA-Pcs4469} Z. G. Wang,  Int. J. Mod. Phys. {\bf A36} (2021) 2150071.

\bibitem{WZG-mole-IJMPA} Z. G. Wang, Int. J. Mod. Phys. {\bf A36} (2021)  2150107.

\bibitem{WZG-XQ-mole-penta} Z. G. Wang and Q. Xin, Chin. Phys. {\bf C45} (2021) 123105.

\bibitem{WZG-XQ-mole-EPJA} Q. Xin and  Z. G. Wang, Eur. Phys. J. {\bf A58} (2022)  110.

\bibitem{WZG-tetra-psedo-NPB} Z. G. Wang and Q. Xin, Nucl. Phys. {\bf B978} (2022) 115761.

\bibitem{WZG-Zcs3985-4110}   Z. G. Wang, Chin. Phys. {\bf C46} (2022) 123106.


\bibitem{WZG-NPB-cucd}  Z. G. Wang,  Nucl. Phys. {\bf B973} (2021) 115592.

\bibitem{WZG-HC-spectrum-PRD}   Z. G. Wang,  Phys. Rev. {\bf D102} (2020)  014018.








\bibitem{X4140-X4700-tetra-ChenHX} H. X. Chen, E. L. Cui, W. Chen, X. Liu and S. L. Zhu, Eur. Phys. J. {\bf C77} (2017)  160.

\bibitem{X4140-X4700-tetra-LuQF} Q. F. Lu and Y. B. Dong, Phys. Rev. {\bf D94} (2016)  074007.


\bibitem{X4140-X4700-tetra-HuangF}  P. P. Shi, F. Huang and W. L. Wang, Phys. Rev. {\bf D103} (2021)  094038.

\bibitem{X4140-X4700-tetra-WuJ}  J. Wu, Y. R. Liu, K. Chen, X. Liu and S. L. Zhu, Phys. Rev. {\bf D94} (2016)  094031.

\bibitem{X4140-X4700-tetra-Ferretti-SB} M. N. Anwar, J. Ferretti and E. Santopinto, Phys. Rev. {\bf D98} (2018)  094015.

\bibitem{X4140-X4700-tetra-HuangHX} X. Liu, H. X. Huang, J. L. Ping, D. Y. Chen and X. M. Zhu, Eur. Phys. J. {\bf C81} (2021)  950.

\bibitem{X4140-X4700-mole-PengFZ} F. Z. Peng, M. J. Yan, M. Sanchez and M. P. Valderrama, Phys. Rev. {\bf D107} (2023)  016001.


\bibitem{X4140-X4700-mole-WangE}  E. Wang, J. J. Xie, L. S. Geng and
E. Oset, Phys. Rev. {\bf D97} (2018) 014017.

\bibitem{X4140-X4700-mole-GuoFK} X. K. Dong, F. K. Guo and B. S. Zou, Progr. Phys. {\bf 41} (2021) 65.

\bibitem{X4140-X4700-cc-ZhongXH} Q. Deng, R. H. Ni, Q. Li and X. H. Zhong,
Phys. Rev. {\bf D110} (2024) 056034.


\bibitem{X4140-X4700-cc-ChenDY} D. Y. Chen, Eur. Phys. J. {\bf C76} (2016)  671.

\bibitem{X4140-X4700-cc-LiuXH}  X. H. Liu, Phys. Lett. {\bf B766} (2017) 117.

\bibitem{cc-Godfrey} T. Barnes, S. Godfrey and E. S. Swanson, Phys. Rev. {\bf D72} (2005) 054026.

\bibitem{WZG-EPJC-P-wave}  Z. G. Wang, Eur. Phys. J. {\bf C79} (2019) 29.

\bibitem{WZG-EPJC-P-2P}  Z. G. Wang,   Chin. Phys. {\bf C48} (2024) 103103.

\bibitem{WZG-HB-spectrum-EPJC} Z. G. Wang, Eur. Phys. J. {\bf C79} (2019)  489.

\bibitem{WZG-tetra-cc-EPJC} Z. G. Wang and Z. H. Yan, Eur. Phys. J. {\bf C78} (2018)  19.

\bibitem{WZG-penta-cc-IJMPA-2050003} Z. G. Wang,  Int. J. Mod. Phys. {\bf A35} (2020)  2050003.

\bibitem{XWWang-penta-mole} X. W. Wang, Z. G. Wang, G. L. Yu and Q. Xin, Sci. China-Phys. Mech. Astron. {\bf65} (2022) 291011.

\bibitem{WangZG-Di-Y4140} Z. G. Wang and Z. Y. Di, Eur. Phys. J. {\bf C79} (2019)  72.

\bibitem{WZG-X4140-X4685} Z. G. Wang, Adv. High Energy Phys. {\bf 2021} (2021) 4426163.

\bibitem{X3915-X4500-EPJA-WZG}   Z. G. Wang,  Eur. Phys. J. {\bf A53} (2017) 19.

\bibitem{X3915-X4500-EPJC-WZG} Z. G. Wang, Eur. Phys. J. {\bf C77} (2017)  78.

\bibitem{WZG-IJMPA-VV} Z. G. Wang, Int. J. Mod. Phys. {\bf A36} (2021) 2150014.


\bibitem{WangZG-Landau} Z. G. Wang,  Phys. Rev. {\bf D101} (2020)  074011.

\bibitem{Wang-Two-particle}  Z. G. Wang, Int. J. Mod. Phys. {\bf A35} (2020) 2050138.

\bibitem{WangZG-Local} Z. G. Wang, arXiv: 2005.12735 [hep-ph].

\bibitem{Reinders85} L. J. Reinders, H. Rubinstein and S. Yazaki, Phys. Rept. {\bf 127} (1985) 1.

\bibitem{Pascual-1984} P. Pascual and R. Tarrach, ``QCD: Renormalization for the practitioner", Springer Berlin Heidelberg (1984).


\bibitem{PDG}  S. Navas et al, Phys. Rev. {\bf D110} (2024) 030001.

\bibitem{Narison-mix} S. Narison and R. Tarrach, Phys. Lett. {\bf B125} (1983) 217.

\bibitem{SVZ79}  M. A. Shifman, A. I. Vainshtein and V. I. Zakharov, Nucl. Phys. {\bf B147} (1979) 385; Nucl. Phys. {\bf B147} (1979) 448.

\bibitem{Colangelo-Review}  P. Colangelo and A. Khodjamirian, hep-ph/0010175.

\bibitem{WangEPJC-1601} Z. G. Wang, Eur. Phys. J. {\bf C76} (2016) 387.

\bibitem{Maiani-II-type} L. Maiani, F. Piccinini, A. D. Polosa and V. Riquer, Phys. Rev. {\bf D89} (2014) 114010.

\bibitem{Nielsen-1401} M. Nielsen and F. S. Navarra,  Mod. Phys. Lett. {\bf  A29} (2014) 1430005.

\bibitem{WZG-XinQ-AAPPS-X3960} Q. Xin, Z. G. Wang and X. S. Yang,
 AAPPS Bull. {\bf 32} (2022)  37.




\end{thebibliography}
\end{document}